\documentclass[aps,prc,twocolumn,nofootinbib,amsmath,showpacs,superscriptaddress,floatfix]{revtex4-1}
\usepackage{graphicx,epsfig,epstopdf,longtable}
\usepackage[dvipsnames]{xcolor}
\usepackage[pdftex,
colorlinks,
pdfpagelabels,
pdfstartview = FitH,
bookmarksopen = true,
bookmarksopenlevel = 2,
bookmarksnumbered = true,
linkcolor = blue!80!black,
plainpages = false,
hypertexnames = false,
urlcolor = blue!80!black,
citecolor = blue!80!black] {hyperref} 
\usepackage[flushleft]{threeparttable}

\usepackage[utf8]{inputenc}

\usepackage{xfrac}

\usepackage{bm}

\usepackage{mathrsfs}
\usepackage{ amssymb }
\usepackage{bm}

\usepackage{natbib} 

\usepackage{nicefrac}
\def\onehalf{\nicefrac{1}{2}}

\usepackage{enumitem}

\usepackage{blindtext}

\usepackage[caption=false]{subfig}
\begin{document}

\title{Restricted spin-range correction in the Oslo Method: The example of nuclear level density and $\gamma$-ray strength function from $^{239}\mathrm{Pu}(\mathrm{d,p}\gamma)^{240}\mathrm{Pu}$}

\author{F. Zeiser}
\email{fabio.zeiser@fys.uio.no}
\affiliation{Department of Physics, University of Oslo, N-0316 Oslo, Norway}

\author{G.M. Tveten}
\affiliation{Department of Physics, University of Oslo, N-0316 Oslo, Norway}

\author{G. Potel}
\affiliation{Facility for Rare Isotope Beams, Michigan State University, East Lansing, MI 48824, USA}

\author{A.C. Larsen}
\affiliation{Department of Physics, University of Oslo, N-0316 Oslo, Norway}

\author{M. Guttormsen}
\affiliation{Department of Physics, University of Oslo, N-0316 Oslo, Norway}

\author{T.A. Laplace}
\affiliation{Department of Nuclear Engineering, University of California, Berkeley, 94720, USA}

\author{S. Siem}
\affiliation{Department of Physics, University of Oslo, N-0316 Oslo, Norway}

\author{D.L. Bleuel}
\affiliation{Lawrence Livermore National Laboratory, Livermore, CA 94551, USA}

\author{B.L. Goldblum}
\affiliation{Department of Nuclear Engineering, University of California, Berkeley, 94720, USA}

\author{L.A. Bernstein}
\affiliation{Department of Nuclear Engineering, University of California, Berkeley, 94720, USA}

\author{F.L. Bello Garrote}
\affiliation{Department of Physics, University of Oslo, N-0316 Oslo, Norway}

\author{L. Crespo Campo}
\affiliation{Department of Physics, University of Oslo, N-0316 Oslo, Norway}

\author{T.K. Eriksen}
\affiliation{Department of Physics, University of Oslo, N-0316 Oslo, Norway}

\author{A. Görgen}
\affiliation{Department of Physics, University of Oslo, N-0316 Oslo, Norway}

\author{K. Hadynska-Klek}
\affiliation{Department of Physics, University of Oslo, N-0316 Oslo, Norway}

\author{V.W. Ingeberg}
\affiliation{Department of Physics, University of Oslo, N-0316 Oslo, Norway}

\author{J.E. Midtbø}
\affiliation{Department of Physics, University of Oslo, N-0316 Oslo, Norway}

\author{E. Sahin}
\affiliation{Department of Physics, University of Oslo, N-0316 Oslo, Norway}

\author{T. Tornyi}
\affiliation{Department of Physics, University of Oslo, N-0316 Oslo, Norway}

\author{A. Voinov}
\affiliation{Department of Physics and Astronomy, Ohio University, Athens, Ohio 45701, USA}

\author{M. Wiedeking}
\affiliation{iThemba LABS, P.O. Box 722, Somerset West 7129, South Africa}

\author{J. Wilson}
\affiliation{Institut de Physique Nucl\'eaire d'Orsay, CNRS/ Univ. Paris-Sud, Universit\'e Paris Saclay, 91406 Orsay Cedex, France}

\date{\today}

\begin{abstract}
The Oslo Method has been applied to particle-$\gamma$ coincidences following the $^{239}\mathrm{Pu}$(d,p) reaction to obtain the nuclear level density (NLD) and $\gamma$-ray strength function ($\gamma$SF) of $^{240}\mathrm{Pu}$. The experiment was conducted with a 12~MeV deuteron beam at the Oslo Cyclotron Laboratory. The low spin transfer of this reaction leads to a spin-parity mismatch between populated and intrinsic levels. This is a challenge for the Oslo Method as it can have a significant impact on the extracted NLD and $\gamma$SF. We have developed an iterative approach to ensure consistent results even for cases with a large spin-parity mismatch, in which we couple Green's Function Transfer calculations of the spin-parity dependent population cross-section to the nuclear decay code RAINIER. The resulting $\gamma$SF shows a pronounced enhancement between 2-4~MeV that is consistent with the location of the low-energy orbital $M1$ scissors mode.
\end{abstract}

\maketitle

\section{Introduction}
\label{sec:int}
Accurate knowledge of neutron induced cross-sections on actinides is important for many applications. From thermal energies up to several MeVs, there is a considerable competition between fission and neutron absorption. This competition, as well as several other factors like the lack of a mono-energetic neutron source in this energy range and the lifetime of short-lived isotopes, pose a challenge for direct cross-sections  measurement. 

Most designs for next generation nuclear reactors are based on fast-neutron induced fission~\cite{Pioro2016}. Therefore, knowledge of the cross-sections for a wider range of incident neutron energies $E_\mathrm{n}$ have become important. In particular, more precise measurements of the $^{239}\mathrm{Pu}$(n,$\gamma$) cross-section below $E_\mathrm{n} \approx 1.5$~MeV are listed as a high priority request by the Nuclear Energy Agency (NEA)~\cite{NEAHPPuNG32}. Calculations for $E_\mathrm{n}$ above the resonance region (i.e., above $\approx 10$\,keV) can be obtained within the statistical Hauser-Feshbach framework~\cite{Hauser1952} and require knowledge of the nuclear level density (NLD) and $\gamma$-ray strength function ($\gamma$SF) of the residual nucleus $^{240}\mathrm{Pu}$. Furthermore, a better knowledge of NLDs and $\gamma$SFs in the actinide region has the potential to improve the nuclear-physics related uncertainties introduced to abundance calculations of heavy-element production in extreme astrophysical environments~\cite{Arnould2007}.

The Oslo method~\cite{Schiller2000, Larsen2011} can be used on particle-$\gamma$ coincidence spectra from transfer reactions to simultaneously extract the NLD and $\gamma$SF below the neutron separation energy $S_\mathrm{n}$. In a campaign to study actinide nuclei, the method has been applied to the compound nuclei $^{231-233}\mathrm{Th}$, $^{232,233}\mathrm{Pa}$, $^{237-239}\mathrm{U}$, $^{238}\mathrm{Np}$~\cite{Guttormsen2012, Guttormsen2013,Guttormsen2014, Tornyi2014a} and  $^{243}\mathrm{Pu}$~\cite{Laplace2015} using different light-ion reactions. So far, all observed NLDs are consistent with a constant temperature~\cite{Ericson1960} level density formula. The $\gamma$SF of these heavy and well-deformed systems show a pronounced enhancement between about 2-4~MeV, which is in the energy range~\cite{Heyde2010} of a low-energy orbital M1 scissors resonance (SR).

The nuclear data community has recently started to take into account these strong M1 SRs, and in two recent studies by Ullmann \textit{et al.}~\cite{Ullmann2014, Ullmann2017}, a significant impact of the SR on the cross-sections calculated for uranium isotopes has been shown. An extraction of the NLD and $\gamma$SF of $^{240}\mathrm{Pu}$ will facilitate similar calculations for $^{239}\mathrm{Pu}(\mathrm{n},\gamma)$. They can be validated by comparison to updated direct measurement by~\citet{Mosby2018} between 10~eV and 1.3~MeV.

\begin{figure*}[tb]
	\centering
	\includegraphics[width=\linewidth]{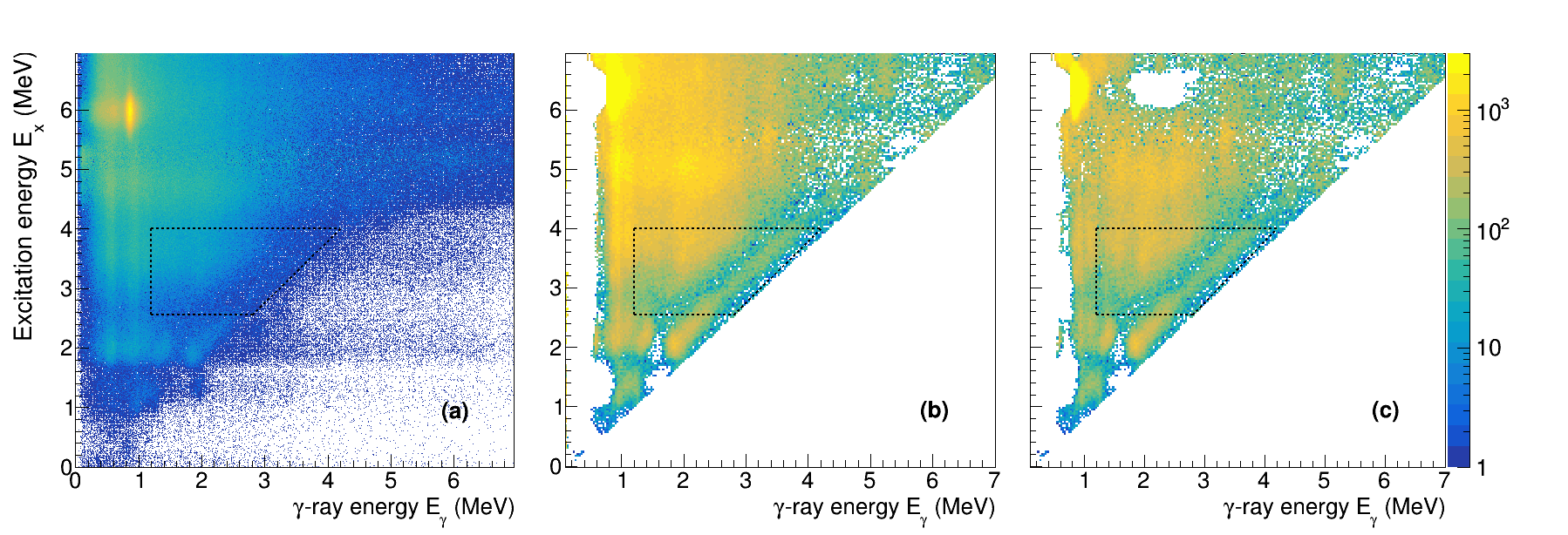}
	\caption{The raw particle-$\gamma$ coincidences for $^{240}\mathrm{Pu}$ (a), the unfolded spectra (b) and the extracted primary-$\gamma$ rays (c). The dotted lines display the region used for the extraction of the NLD and $\gamma$SF. Before unfolding, all events with $E_\gamma > E_\mathrm{x} + \delta E_\gamma$ have been removed as they only represent noise, where $\delta E_\gamma$ is the detector resolution.}
	\label{fig:m_alfna_and_fg}
\end{figure*}

Larsen \textit{et al.}~\cite{Larsen2011} have shown that the population of a limited spin range make it necessary to correct the slopes of $\gamma$SFs extracted with the Oslo Method. In the previous experimental studies on actinides~\cite{Guttormsen2012,  Guttormsen2013, Guttormsen2014, Tornyi2014a, Laplace2015}, first indications of the impact of a low spin transfer using the (d,p) reaction mechanism were observed and an improvised procedure for the correction was developed. More recently, we have presented a systematic analysis of the effect of a realistic spin distribution on both the NLD and $\gamma$SF for the $^{239}\mathrm{Pu}(\mathrm{n},\gamma)^{240}\mathrm{Pu}$ reaction \cite{Zeiser2018a}.

In this article, we will present the NLD and $\gamma$SF of $^{240}\mathrm{Pu}$ analyzed with the Oslo Method. We develop an iterative procedure to correct for the bias introduced in the Oslo Method for (d,p) reactions on heavy nuclei due to a spin-parity population mismatch.

\section{Experimental methods and data analysis}
\label{sec:exp}

The $^{239}\mathrm{Pu}(\mathrm{d},\mathrm{p})^{240}\mathrm{Pu}$ experiment was conducted using a 12~MeV deuteron beam extracted from the MC-35 Scanditronix Cyclotron at the Oslo Cyclotron Laboratory (OCL). The $0.4\,\mathrm{mg}/\mathrm{cm}^2$ thick $^{239}\mathrm{Pu}$ target was purified by an anion-exchange resin column procedure~\cite{Henderson2011} prior to electroplating it onto a $2.3\,\mathrm{mg}/\mathrm{cm}^2$ beryllium backing. A $\gamma$-ray assay of the resulting target revealed the $^{239}\mathrm{Pu}$ purity to be $>99.9$\%. \newline

Particle-$\gamma$ coincidences were measured with the SiRi particle telescopes~\cite{Guttormsen2011} and CACTUS $\gamma$-ray detector array~\cite{Guttormsen1990}. SiRi consists of 64 silicon particle telescopes with a thickness of 130 $\mu$m for the front ($\Delta E$) and 1550 $\mu$m for the back ($E$) detectors. In this experiment they were placed in a backward position with respect to the beam direction, covering azimuthal angles from $126^\circ$ to $140^\circ$. Compared to the forward direction, this configuration reduces the contribution of elastically scattered deuterons and populates a broader and higher spin-range.
The CACTUS array was composed of 26 lead collimated $5^{\prime\prime} \times 5^{\prime\prime}$ NaI(Tl) crystals with a total efficiency of 14.1(2)$\%$ at $E_\gamma= 1.33~\mathrm{MeV}$ (measured with a $^{60}\mathrm{Co}$ source) that surrounded the target chamber and the particle telescopes. Additionally, four Parallel Plate Avalanche Counters (PPAC)~\cite{Tornyi2014} were used to detect fission events.
The back detectors of SiRi were used as master gates for a time-to-digital converter (TDC). The NaI(Tl) detectors were delayed by $\approx 400\,\mathrm{ns}$ and individually served as stop signals. The signals were processed by a leading edge discriminator and the resulting time walk was corrected for by the procedure given in Ref.~\cite{Guttormsen2011}. The prompt particle-$\gamma$ coincidences were sorted event-by-event from a 28\,ns wide time-window and the background from random coincidences was subtracted. The amount of deposited energy depends on the outgoing particle type, which facilitated the selection of (d,p) events by setting proper gates in a $\Delta E$-$E$ matrix. The spectra were calibrated using reaction kinematics, which also allowed translation of the deposited particle energy to the initial excitation energy $E_\mathrm{x}$ of the residual nucleus $^{240}\mathrm{Pu}$. The $\gamma$-ray spectra for each excitation energy $E_\mathrm{x}$ were unfolded following the procedure of Ref.~\cite{Guttormsen1996}, however using new response functions measured in 2012~\cite{Campo2016}. In this work we used the Oslo method software v1.1.2 \cite{Oslo-v1.1.2}.

To select the $\gamma$ decay channel, only excitation energies $E_\mathrm{x}$ below the neutron separation energy ($S_n = 6.534~\mathrm{MeV}$~\cite{Singh2008}) were considered. The energy range was further constrained by pile-up of $\gamma$-rays and the onset of fission events at $E_\mathrm{x} \approx 4.5$~MeV. The latter was previously identified as sub-barrier fission~\cite{Back1974, Hunyadi2001}. A more detailed analysis of the prompt fission $\gamma$-rays can be found in Ref.~\cite{Rose2017}. The final extraction regions were $E_{\gamma}^\mathrm{min}=1.2~\mathrm{MeV}$, $E_{\mathrm{x}}^\mathrm{min}=2.5~\mathrm{MeV}$, $E_{\mathrm{x}}^\mathrm{max}=4.0~\mathrm{MeV}$.

We applied an iterative subtraction technique to obtain the energy spectrum of the primary (also called first generation) $\gamma$-rays from the initial spectrum, which includes all $\gamma$ decay cascades. The principal assumption of the first-generation method~\cite{Guttormsen1987} is that the $\gamma$ decay from any excited state is independent of its formation. The branching ratio is an inherent property of a state. Thus, the assumption is automatically fulfilled if levels have the same probability to be populated by the decay of higher-lying states as directly by nuclear reactions (e.g. via the (d,p) reaction). As we consider the quasi-continuum, we can relax the strict conditions and apply statistical considerations so we only require that in a given excitation energy bin all levels with the same spin-parity are populated approximately equally (instead of specific states). In addition, the population probability of levels with a given spin-parities should be approximately constant as a function of the excitation energy. In Sec. \ref{sec:method} we will show that this condition is not satisfied and we propose a procedure to minimize the impact of the violation of this assumption. For a thorough discussion of other possible errors and uncertainties in this method, see Ref. \cite{Larsen2011}. The coincidence matrices are displayed in Fig. \ref{fig:m_alfna_and_fg}.

\section{Extraction of NLD and $\gamma$SF}
According to Fermi's golden rule, the decay rate from an initial state to a final state can be decomposed into the transition matrix element and the level density $\rho(E_\mathrm{f})$ at the final state $E_\mathrm{f} = E_i - E_\gamma$ \cite{Dirac1927,Fermi1950}. In the regime of statistical $\gamma$-rays, we  consider ensembles of initial and final states, thus probing decay properties averaged over many levels. We assume that any decay mode can be build on the ground and excited state in the same way, i.e. there is no spin-parity or excitation energy dependence, which is a generalized version of the Brink-Axel hypothesis~\cite{Brink1955,Axel1962}. Thus, the decay properties do not depend on the specific levels, but only on the energy difference between them. Consequently, the dependence on initial and final states is reduced to a single dependence on the energy difference given by the $\gamma$-ray energy $E_\gamma$. The decay probability corresponding to the first-generation matrix $P(E_i,E_\gamma)$ can therefore be factorized into the level density of the final excitation energy $\rho(E_\mathrm{f})$ and the transmission coefficient $\mathscr{T}(E_\gamma)$~\cite{Schiller2000}:
\begin{equation}
P(E_i,E_\gamma) \propto \rho(E_\mathrm{f}) \mathscr{T}(E_\gamma).
\label{eq:LevelDensityTransmissionCoefficient}
\end{equation}
The validity of the Brink-Axel hypothesis in the quasi-continuum has recently been shown for several nuclei \cite{Larsen2017, Campo2018}, amongst them the actinide nucleus $^{238}\mathrm{Np}$~\cite{Guttormsen2016}. The level density  $\rho(E_\mathrm{f})$ and transmission coefficient $\mathscr{T}(E_\gamma)$ were obtained by a fit to $P(E_i,E_\gamma)$~\cite{Schiller2000}. Note that this procedure does not require any initial assumptions on the functional form of $\rho$ and $\mathscr{T}$. However, any transformation $\tilde{\rho}$ and $\tilde{\mathscr{T}}$ with the parameters $\alpha$, $A$ and $B$ gives identical fits to the matrix $P(E_i,E_\gamma)$:~\cite{Schiller2000}
\begin{align}
\tilde{\rho} (E_i - E_\gamma) & = A \exp[\alpha \,(E_i - E_\gamma) ]\,  \rho(E_i - E_\gamma), \label{eq:rho_trans}\\
\tilde{\mathscr{T}}(E_\gamma) &= B \exp[ \alpha \, E_\gamma ]\, \mathscr{T}(E_\gamma). \label{eq:t_trans}
\end{align}
The determination of the transformation parameters corresponding to the correct physical solution, i.e. the normalization of the NLD and $\gamma$SF, is discussed in the next section.

\section{Initial extraction of the level density and transmission coefficient}
\label{sec:exp_extraction}
For the normalization of the level density, $\rho$, we need at least two reference points, such that we can determine the parameters $A$ and $\alpha$ in Eq. \eqref{eq:rho_trans}. At low excitation energies, our data are matched to discrete levels~\cite{NNDC} up to the critical energy $E_\mathrm{crit}\approx 1.3~\mathrm{MeV}$ where we expect the low-lying level scheme to be complete. At the neutron separation energy $S_\mathrm{n}$, we calculate $\rho(S_n)$ under the assumption of equal parity distribution from the average neutron resonance spacing for s-waves, $D_0$, taken from RIPL-3~\cite{RIPL3} following Ref.~\cite{Schiller2000}:
\begin{align}
&\rho(S_n)  \label{eq:rhoFromD0} \\
&= \frac{2 \sigma^2}{D_0} \frac{1}{(J_t+1) \exp[-(J_t+1)^2 / 2\sigma^2] + J_t \exp[-J_t^2 / 2 \sigma^2]}. \nonumber
\end{align}
Here $J_t$ is the ground-state spin of the target nucleus $^{239}\mathrm{Pu}$.

\begin{figure}[tb]
	\centering
	\includegraphics[width=\linewidth]{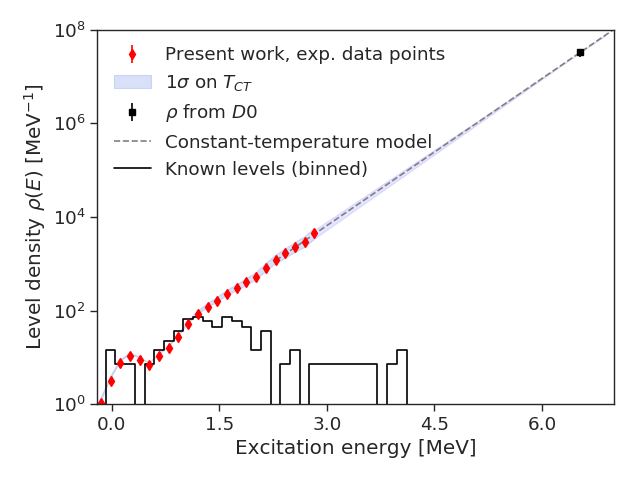}
	\caption{Initial analysis of the total NLD for $^{240}\mathrm{Pu}$. The NLD is normalized to the discrete levels (in 140 keV bins)~\cite{NNDC} at low excitation energies and to $\rho(S_n)$ calculated from $D_0$~\cite{RIPL3}, using a constant temperature interpolation with $T_{\mathrm{CT}}=0.415(10)$.}
	\label{fig:counting}
\end{figure}

\begin{table*}
	\centering
	\caption{Parameters used to extract the initial level density and $\gamma$-strength function (see text).}
	\begin{threeparttable}
		\centering
		\begin{tabular}{l*{10}{c}} \toprule
			$S_n$ & $a$ & $E_1$ & $\sigma(S_n)$ & $D_0$ & $\rho(S_n)$ &$T_\mathrm{CT}$ & $\langle \Gamma_\gamma (S_n)\rangle$ \\
			$[\mathrm{MeV}]$ & $[\mathrm{MeV}^{-1}]$ & $[\mathrm{MeV}]$ &  & $[\mathrm{eV}]$ &  $[10^6 \,\mathrm{MeV}^{-1}]$ & $[\mathrm{MeV}]$ & $[\mathrm{MeV}]$ \\ \colrule
			6.53420(23)\tnote{a} ~& 25.16(20)\tnote{b}~ & 0.12(8)\tnote{b}~ & 8.43(80)\tnote{d}~ & 2.20(9)\tnote{c}~ & 32.7(66) & 0.415(10) & 43(4)\tnote{c} \\
			\botrule
		\end{tabular}
		\begin{tablenotes}[flushleft, para]
			\item[a] Ref.~\cite{Singh2008}
			\item[b] Ref.~\cite{Egidy2005}
			\item[c] Ref.~\cite{RIPL3}
			\item[d] Assuming a 10\% uncertainty
		\end{tablenotes}
	\end{threeparttable}
	\label{tab:ExtractionParameters}
\end{table*}

We use the spin distribution $g(E_\mathrm{x},I)$ proposed by \citet[Eq. (3.29)]{Ericson1960}\footnote{The same spin distribution is often attributed to the subsequent work of \citet{Gilbert1965}.} together with the rigid-body moment of inertia approach for the spin cut-off parameter $\sigma$ from 2005 by~\citet{Egidy2005}:
 \begin{align}
 	g(E_\mathrm{x},I) & = \frac{2 I +1}{2 \sigma^2(E_\mathrm{x})} \exp[-(I+1/2)^2 / 2 \sigma^2], \label{eq:SpinDistGC}\\
 	\sigma^2(E_\mathrm{x}) & = 0.0146 A^{2/3} \frac{1+ \sqrt{4 a U(E_\mathrm{x})}}{2a}, \label{eq:SpinCutoffEB06}
 \end{align}
 where $A$ is the mass number of the nucleus, $a$ is the level density parameter, $U(E_\mathrm{x})=E_\mathrm{x} - E_1$ is the intrinsic excitation energy, and $E_1$ is the back-shift parameter. All parameters are listed in Tab. \ref{tab:ExtractionParameters}

Since there is a gap of approximately 3.5~MeV between the highest excitation energy of the extracted level densities and the neutron separation energy $S_n$, an interpolation is used to connect the datasets. In accordance with the findings for other actinides~\cite{Guttormsen2013}, we use the constant temperature (CT) level density formula~\cite{Ericson1960}
\begin{align}
\rho_\mathrm{CT}(E_\mathrm{x}) = \frac{1}{T_{\mathrm{CT}}} \exp \frac{E_\mathrm{x} - E_0}{T_{\mathrm{CT}}},
\label{eq:constTemp}
\end{align}
with the shift in excitation energy $E_0$ given by
\begin{align}
E_0 = S_n - T_{\mathrm{CT}} \ln [\rho(S_n)T_{\mathrm{CT}}].
\end{align}
The best fit is obtained for a constant temperature of $T_{\mathrm{CT}}=0.415(10)~\mathrm{MeV}$. Only a limited number of data points are available for the fit which are well above $E_\mathrm{crit}$. This makes a proper interpretation of the uncertainty on the fit parameters difficult. This is the main contribution to the systematic error, which is shown as an error band in the results in Fig. \ref{fig:counting}.

For the transmission coefficient $\mathscr{T}$, the remaining parameter $B$ is determined by normalization to the average total radiative width $\langle \Gamma_\gamma (S_\mathrm{n})\rangle$ from (n,$\gamma$) experiments~\cite{RIPL3}, under the assumption of equal parity using~\cite{Voinov2001, Kopecky1990}
\begin{align}
& \langle \Gamma_\gamma (S_n,J_t \pm \onehalf, \pi_t)\rangle \notag\\  =  & \quad \frac{B}{4 \pi \rho(S_n,J_t \pm \onehalf,\pi_t)}  \int_0^{S_n} dE_\gamma\, \mathscr{T}(E_\gamma)
	\rho(S_n-E_\gamma) \notag\\
& \quad \times \sum_{j=-1}^{1} g(S_n-E_\gamma, J_t \pm \onehalf + j),
	\label{eq:GammaGamma}
\end{align}
where $\pi_t$ is the ground-state parity of the target nucleus $^{239}\mathrm{Pu}$. Note that the sum in Eq. \eqref{eq:GammaGamma} runs over all available final states of $^{240}\mathrm{Pu}$, where we consider only spins $J_t \pm \onehalf + j$ that can be reached by one primary dipole transition after neutron capture, i.e. $j = -1, 0, 1$.
The $\gamma$-ray strength function $f$ is obtained under the same assumption of a dominance of dipole strength, $L$=1, so $f \simeq f_{E1} + f_{M1}$, and 
\begin{align}
	f(E_\gamma) = \frac{\mathscr{T}(E_\gamma)}{2 \pi E^{2L+1}} \simeq \frac{\mathscr{T}(E_\gamma)}{2 \pi E_\gamma^{3}}.
\end{align}
To specify the integral in Eq. \eqref{eq:GammaGamma} completely, we use a log-linear extrapolation in the $\gamma$SF below $E_\gamma^\mathrm{min}$ and a log-linear extrapolation in $\mathscr{T}$ between $E_\gamma^\mathrm{max}$ and $S_\mathrm{n}$.

\section{Corrections due to spin-parity mismatch}
\label{sec:method}
First indications that a limited spin-range of the levels populated in a given reaction has an impact on the Oslo method have been discussed in Ref.~\cite{Larsen2011}. Due to the low angular momentum transfer expected for light-ion reactions, and in particular the (d,p) transfer reaction, the higher spin states that are already available at $E_\mathrm{x} \approx 2~\mathrm{to}~6$~MeV in heavy nuclei may not be populated. In Ref.~\cite{Guttormsen2012} an ad-hoc method was developed to correct for observations that were attributed to the limited angular momentum transfer. This correction has subsequently been applied to other heavy nuclei~\cite{Guttormsen2013, Guttormsen2014, Tornyi2014a, Laplace2015, Giacoppo2015}. In a recent analysis on systematic errors for (d,p)$^{240}\mathrm{Pu}$ we have demonstrated that the application of the Oslo Method produces consistent results when the spin-parity dependent population probability $g_\mathrm{pop}$ equals the theoretically expected distribution of the intrinsic levels $g_\mathrm{int}$. However, when there is a large mismatch in the spin-parity distributions we have also shown that the aforementioned ad-hoc method lead to significant distortions in the NLD and $\gamma$SF~\cite{Zeiser2018a}. We will denote the extracted quantities as the \textit{apparent} NLD and $\gamma$SF, and distinguish them from the \textit{true} NLD and $\gamma$SF that would have been observed with an ideal, bias-free method. In absence of an ideal method, our goal is to find a consistent set of NLD and $\gamma$SF, where we define consistency as follows: if we provide this set as input to a nuclear decay code like RAINIER~\cite{L.E.Kirsch2019}, the generated synthetic data should match the experimentally obtained coincidences. This grantees at the same time that the analysis of the synthetic data yields the same \textit{apparent} NLD and $\gamma$SF as those determined from the \textit{naive}\footnote{In the sense that the experimental analysis does not inherently take into account a spin-parity mismatch.} experimental analysis. In this section we extend the analysis of Ref.~\cite{Zeiser2018a} in order to retrieve a consistent set of NLD and $\gamma$SF for $^{240}\mathrm{Pu}$ for the same reaction. This approach is, however, easily generalizable to other target nuclei and in principle also applicable for other light-ion reactions. 

We will start with a brief overview of the procedure and then discuss each step in more detail:
\begin{enumerate}[itemsep=0pt]
	\item Calculate the spin-parity distribution of the population probability $g_\mathrm{pop}$, and the distribution of the intrinsic levels $g_\mathrm{int}$ for each excitation energy bin $E_x$.
	\item Generate a synthetic coincidence dataset for an artificial nucleus resembling $^{240}\mathrm{Pu}$, given the spin distributions, and the trial NLD and $\gamma$SF.
	\item Analyze and compare the \textit{apparent} NLD and $\gamma$SF from the synthetic dataset and experimental coincidences using the Oslo Method. 
	\item Adjust the trial NLD and $\gamma$SF and repeat step 2 and 3. Adopt the solution with the smallest difference between experimental and synthetic coincidence spectra.
\end{enumerate}

\begin{figure}[tb]
	\centering
	\includegraphics[width=1.\linewidth,
					trim={2mm 2mm 5mm 0mm},clip]{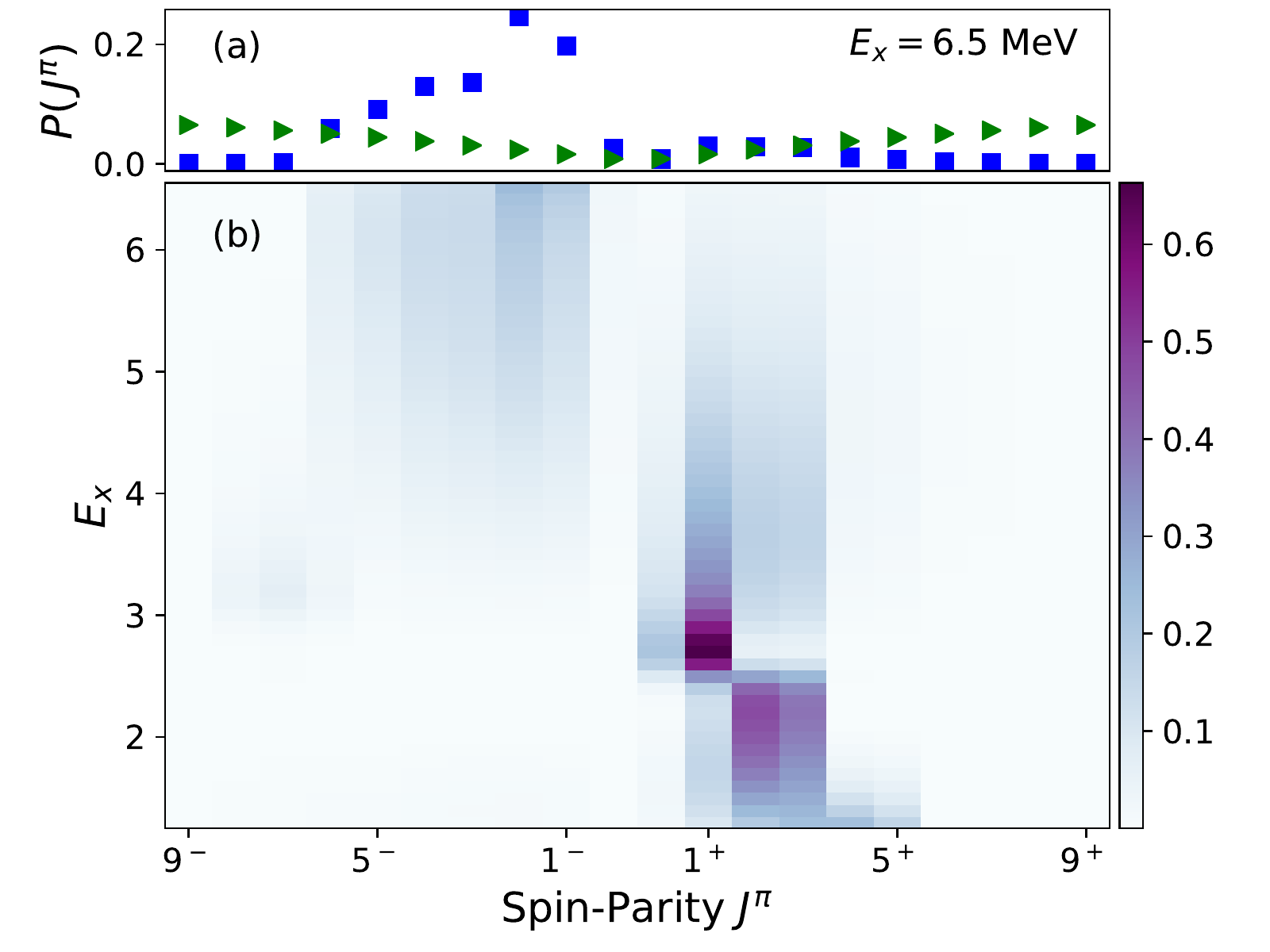}
	\caption{Population probability $g_\mathrm{pop}(E_\mathrm{x},J,\pi)$ of levels in the $^{239}\mathrm{Pu}\mathrm{(d,p)}^{240}\mathrm{Pu}$ reaction as a function of excitation energy $E_x$, and spin-parity $J^\pi$. (a) Projection of $g_\mathrm{pop}(E_\mathrm{x})$ (blue squares) for the highest excitation energy, $E_\mathrm{x}=6.5~\mathrm{MeV}$, which reveals a strong asymmetry in the populated parities. We observe that $g_\mathrm{int}$ (green triangles) is much broader then $g_\mathrm{pop}$ (blue squares). Note that the distributions are normalized to 1 summing over all $J^\pi$ in each $E_x$ bin, but the plot ranges only between $J^\pi = 9^\pm$. }
	\label{fig:popToTot}
\end{figure}

To calculate the population probability $g_\mathrm{pop}$ for each $J^\pi$ in the residual nucleus following a (d,p) reaction, we have to distinguish between two reaction mechanisms. First, we consider direct processes, i.e. the breakup of a deuteron with emission of a proton, followed by the formation of a compound nucleus with the remaining neutron and the target. Spin-parity dependent cross section are calculated for the angles covered in the experiment within the Green's Function Transfer formalism described in Ref.~\cite{Potel2015, Potel2017}. The neutron-nucleus interactions are modeled by the dispersive optical model potential (OMP) of \citet{Capote2008} implemented through potential no. 2408 listed in RIPL-3 \cite{RIPL3}. The usage of a dispersive OMP improves the predictive power for $E_\mathrm{x} < S_\mathrm{n}$. Note that we did not use the OMP in the context of full coupled-channels calculations, which would have explicitly accounted for the coupling to rotational states. We expect that this will lead to an underestimation of the absorption cross-section of about 20\%; however the relative population of the different spins and parities should essentially be unaffected. We normalize the population cross-sections to 1 for each $E_\mathrm{x}$ bin, thus obtaining the probability distribution $g_\mathrm{pop}$. Figure \ref{fig:popToTot} shows the results for the population spin-parity distribution $g_\mathrm{pop}(E_\mathrm{x},J,\pi)$.


Compound reactions are the second mechanism leading to $^{240}\mathrm{Pu}$ as a residual nucleus: proton evaporation after fusion of the deuteron and target nucleus and the inelastic excitation of the target to energies above the proton emission threshold. The spin-parity integrated cross-section for these processes has been estimated to be $\approx 0.5\,\mathrm{mb}/(\mathrm{MeV}\, \mathrm{sr})$ using the statistical framework of the TALYS nuclear reactions code v1.8~\cite{Koning2007}. This is an order of magnitude smaller than for the direct process and therefore neglected. The low cross-sections are reasonable as the deuteron beam energy of 12~MeV is below the Coulomb barrier of about 14.46~MeV, where the latter is calculated with a radius parameter $r_0=1.26$\,fm~\cite{Soukhovitskii2004}.

\begin{figure*}[t]
	\centering
	\subfloat{%
		\includegraphics[width=0.99\columnwidth]{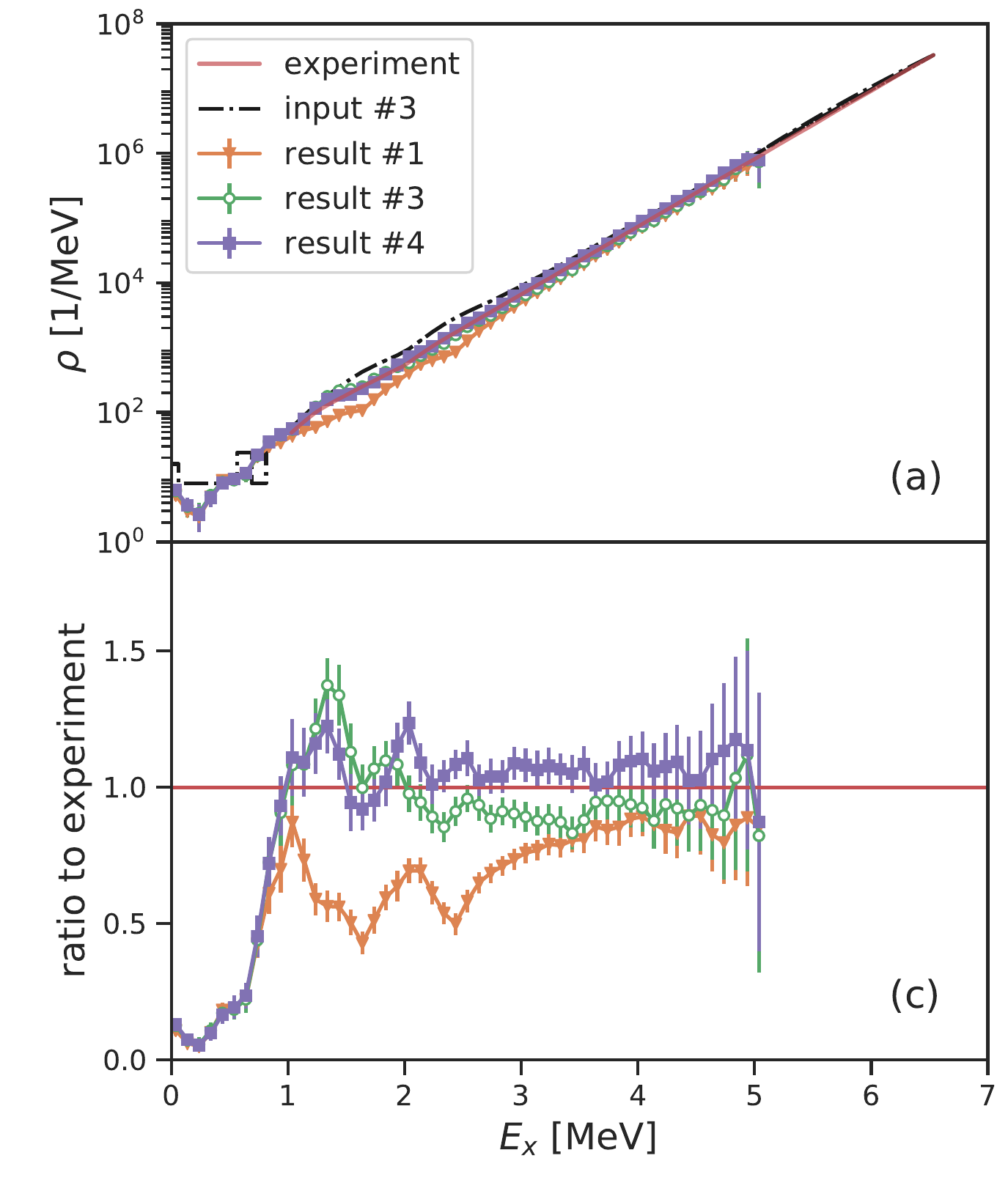}%
		\label{plot:nld_analyzed}%
	}\qquad
	\subfloat{%
		\includegraphics[width=0.99\columnwidth]{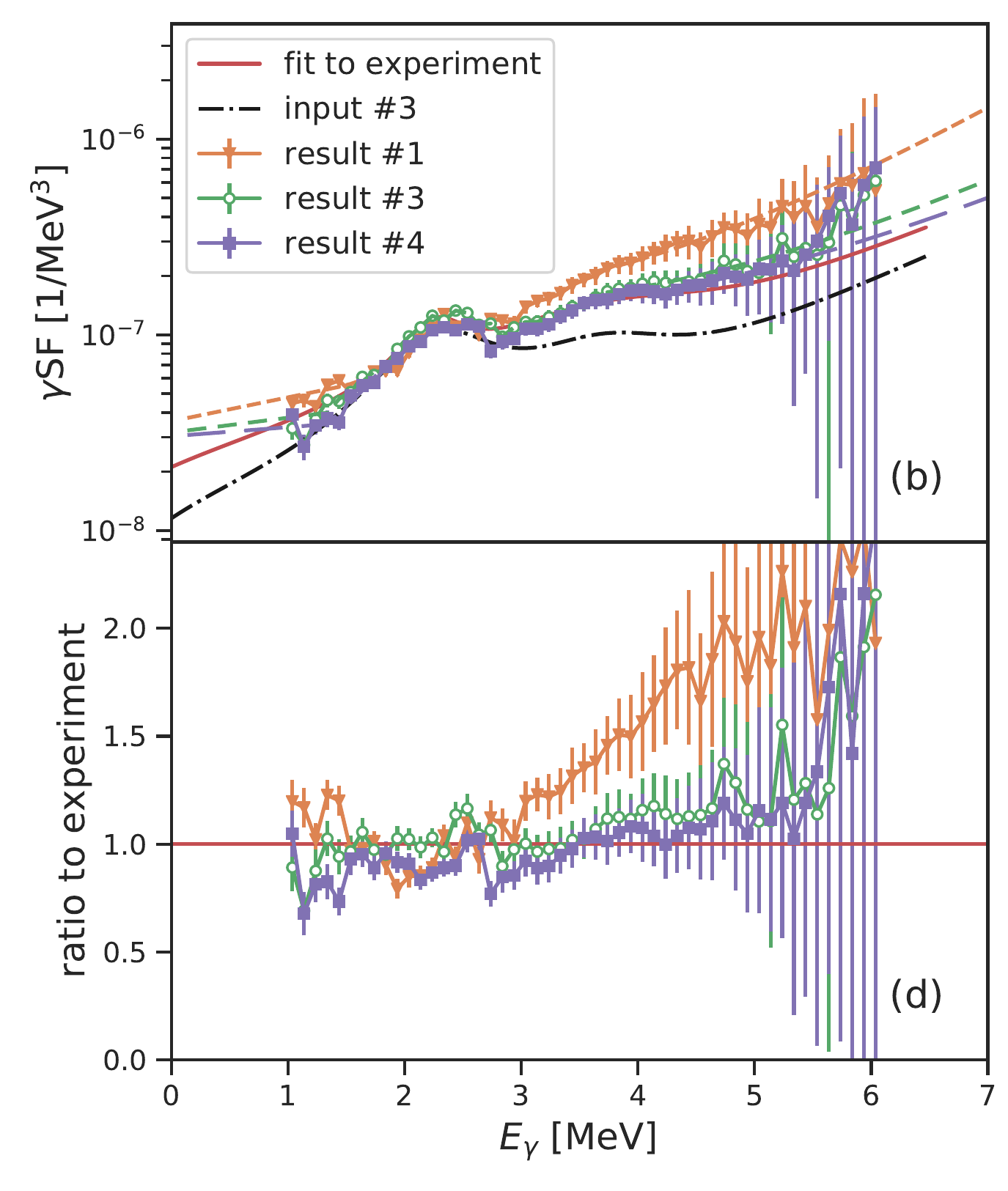}%
		\label{plot:gsf_analyzed}%
	}
	\caption{Upper panels: NLD (a) and $\gamma$SF (b) extracted with the Oslo method from synthetic data (iteration 1, 3 and 4) compared to those extracted from the experimental coincidence data in Sec. \ref{sec:exp_extraction}. The $\gamma$SFs are compared to the fit of the experimental data points. As a guide to the eye, the data are connected by solid lines and dashed lines denote the extrapolations assumed for the Oslo Method. Lower panels: Ratios of the NLD (c) and $\gamma$SF (d) extracted from synthetic data to those from the experimental coincidence data. The error-bars are a combination of statistical and proposed systematic error (mostly due to potential non-statistical decay at high $E_\gamma$) as retrieved from the Oslo Method when analyzing the synthetic data. Note that the analysis of synthetic data created from iteration 3 (input is displayed) results in a NLD and $\gamma$SF that closely resemble the experimental analysis.}
	\label{fig:nld_gsf_RAINIER}
\end{figure*}

To study the effect on the extracted NLD and $\gamma$SF, we generate a synthetic dataset with the statistical nuclear decay code RAINIER v1.4.1~\cite{Kirsch2018, L.E.Kirsch2019}. This code uses a Monte Carlo approach to generate levels of an artificial nucleus and simulate $\gamma$-emission cascades via $E1$, $M1$, or $E2$ transitions. The analysis library facilitates the extraction of the $\gamma$-ray spectra (first or all generations) emitted from each initial excitation energy bin $E_x$. The matrix including the $\gamma$-ray spectra of all generations substitutes for the experimental particle-$\gamma$ coincidence in the further analysis. The input parameters have been chosen to resemble the $^{240}\mathrm{Pu}$ nucleus and the analysis in the previous section. The initial settings are summarized below, and a comprehensive list including the analysis code can be found online\footnote{\url{https://github.com/fzeiser/240Pu_article_supplement}}:
\begin{itemize}[itemsep=0pt]
	\item Discrete levels up to 1.037 MeV (18 levels).
	\item Above 1.037~MeV: Generated levels from the NLD extracted in Sec. \ref{sec:exp_extraction} with the nearest neighbor spacing according to the Wigner distribution \cite{Wigner1957}.
	\item Intrinsic spin distribution $g_\mathrm{int}(E_\mathrm{x},J)$ following Eq. \eqref{eq:SpinDistGC}, with a spin-cut parameter $\sigma$ of Eq. \eqref{eq:SpinCutoffEB06} (assumes equiparity).
	\item Spin-parity dependent population probabilities  $g_\mathrm{pop}(J,\pi)$ from our calculations. \footnote{Note that we did not include the excitation energy dependence of the population cross-section for this analysis, although it could in principle be included to give a more stringent test of the first generation method.}
	\item $\gamma$SF as extracted in Sec. \ref{sec:exp_extraction}, fitted by two $E1$ constant temperature Generalized Lorentzians (GLO)~\cite{Kopecky1990}, two $M1$ Standard Lorentzians (SLO), and including Porter-Thomas fluctuations~\cite{Porter1956}. The $E2$ component was assumed to be negligible.
	\item Internal conversion model: BrIcc Frozen Orbital approximation~\cite{Kibedi2008}.
\end{itemize}

Due to the strong parity dependence of $g_\mathrm{pop}$, the generated simulated coincidence spectra depend on the decomposition of the $\gamma$SF into its $E1$ and $M1$ components. We performed a $\chi^2$ fit of the centroid, the peak cross section and width of each resonance of the $\gamma$SF  simultaneously using the differential evolution algorithm by \citet{Storn1997}. In addition to our data $\bm{Y_\mathrm{sum}}$, which measures only the summed $\gamma$SF ($M1$+$E1$), we include the data $\bm{Y_\mathrm{E1/M1}}$ of Kopecky \textit{et al.} \cite{Kopecky2017, Chrien1985} around $S_n$, which resolve the $E1$ and $M1$ components. There are no measurements for the Giant Dipole Resonance (GDR) of $^{240}\mathrm{Pu}$. However, as the GDR is expected to vary little between the plutonium isotopes, we also include $^{239}\mathrm{Pu}(\gamma,\mathrm{abs})$ measurements (again included in $\bm{Y_\mathrm{sum}}$) by \citet{Moraes1993} and \citet{Gurevich1976}. A third dataset by \citet{Berman1986} is yield systematically lower cross-sections than the first two measurements, which where consistent within the error-bars. Therefore we did not include the data of \citet{Berman1986} in the fit. Each term is weighted by the experimental uncertainty of the  datapoint. The total $\chi^2$ is then given as the sum over the $\chi^2$s for the summing data $\bm{Y_\mathrm{sum}}$ ($E1+M1$) and data $\bm{Y_\mathrm{E1/M1}}$ that resolve the $M1$ and $E1$ contributions:
\begin{align}
\chi^2= \sum_{i \in \bm{Y_\mathrm{sum}}} \chi^2_\mathrm{sum}
+\sum_{i \in \bm{Y_{E1}}} \chi^2_{E1}
+\sum_{i \in \bm{Y_{M1}}} \chi^2_{M1}.
\label{eq:Chi2}
\end{align}


The generated coincidence data are analyzed with the Oslo Method and the results are displayed in Fig. \ref{fig:nld_gsf_RAINIER}. We can quantify how consistent the input NLD and $\gamma$SF are by construction of the ratio $r$ of the apparent NLD and $\gamma$SF analyzed from synthetic data to the experimental analysis (see Sec. \ref{sec:exp_extraction}). We extract this ratio for each iteration. For the NLD this means that below 3~MeV we compare to the data points, whereas above 3~MeV we use the CT extrapolation. In case of the $\gamma$SF, we compare to its fit, so the sum of the 2 GLOs and 2 SLOs. The inverse of the ratio $r$ is used as a bin-by-bin correction $z= (1/r) - 1$ to the input NLD and $\gamma$SF of iteration $n$, such that we generate the input for the next iteration, $n+1$: 
\begin{align}
I_{n+1}=I_n \left(1+ \frac{1}{2} z\right),
\end{align}
where $I$ is the input NLD or $\gamma$SF, respectively. We introduced an additional factor of $1/2$, which can be seen as reduction of the step-size of the correction $z$. This increased the stability of the solution. As an example, looking at the first iteration, we find that the analyzed NLD from the synthetic data at 2.5~MeV is only 50\% of the experimentally observed NLD. We would therefore increase the input NLD for the next iteration by 25\% in this bin (and process all other bins of the NLD and $\gamma$SF in the same manner). For the first iterations we observe that the changes impact $\langle \Gamma_\gamma\rangle$ by about 25\%. As $\langle \Gamma_\gamma \rangle_\mathrm{exp}$ is determined from independent measurements, we enforce a match by rescaling the predicted input $\gamma$SF. 

After only 3 to 4 iterations, we observe that the $\gamma$SF and NLD have approximate converged, with the exception of the higher energy region of the $\gamma$SF. The reproduction of the $\gamma$SF above $E_{\mathrm{\gamma}}^\mathrm{max}=4.0~\mathrm{MeV}$ remains challenging. The corresponding fit region in the first-generation matrix is formed by non-statistical decays, thus it is not obvious that the Oslo Method should be applicable in this regime. In addition, the comparison in this regime is sensitive to the choice of the extrapolation of the initial $\gamma$SF.

\begin{figure}[tb]
	\centering
	\includegraphics[width=1.\linewidth,
	trim={0mm 0mm 15mm 10mm},clip]{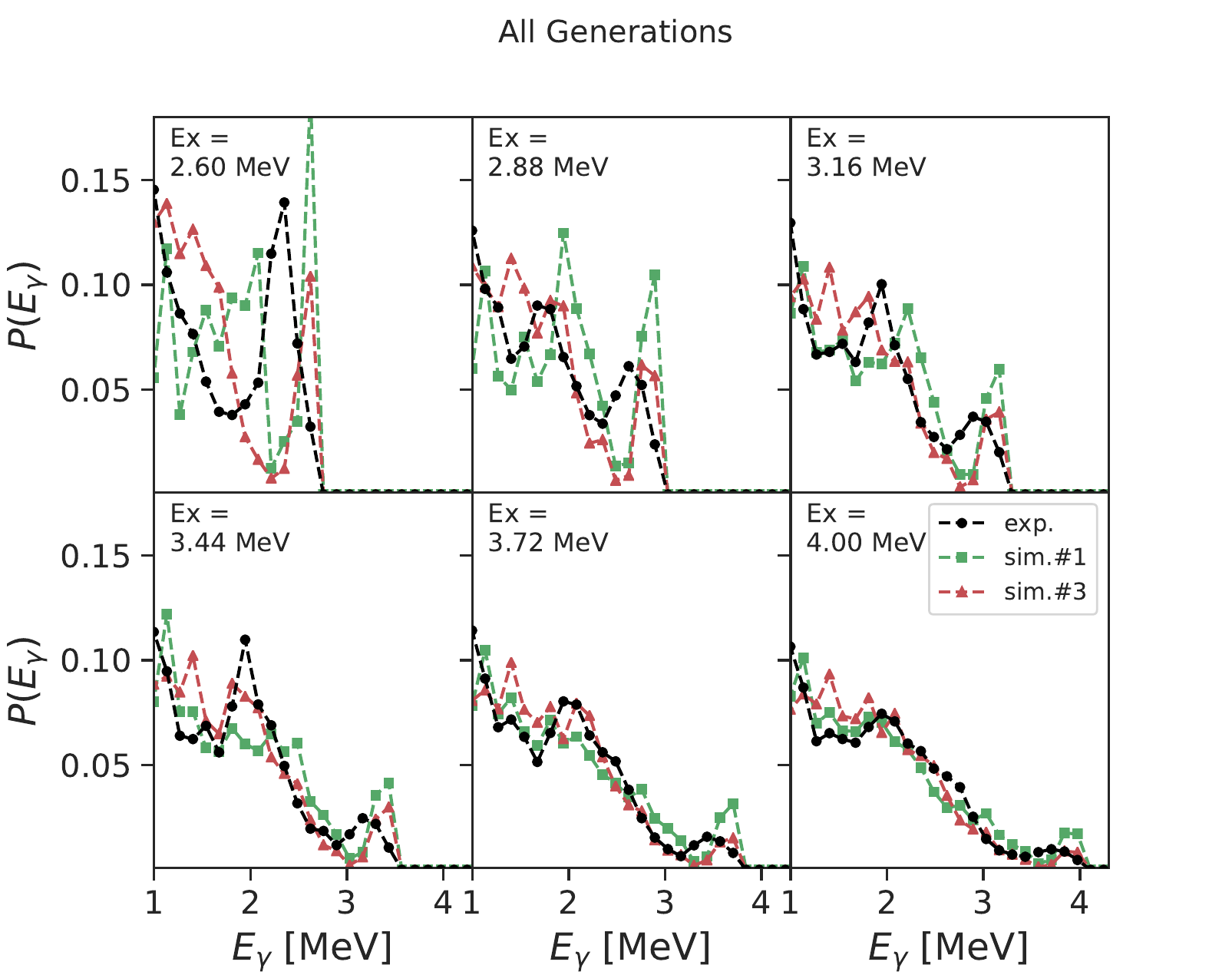}
	\caption{Comparison of experimental coincidence data with the synthetic data from iteration 1 and 3. In general, the results of iteration 3 match the data quite well, but they fail to reproduce the spectra for the lowest excitation energy bin. The comparison region was chosen in accordance with the extraction region specified in Sec. \ref{sec:exp}.}
	\label{fig:spectra}
\end{figure}

\begin{figure}[tb]
	\centering
	\includegraphics[width=1.\linewidth,
	trim={0mm 0mm 15mm 10mm},clip]{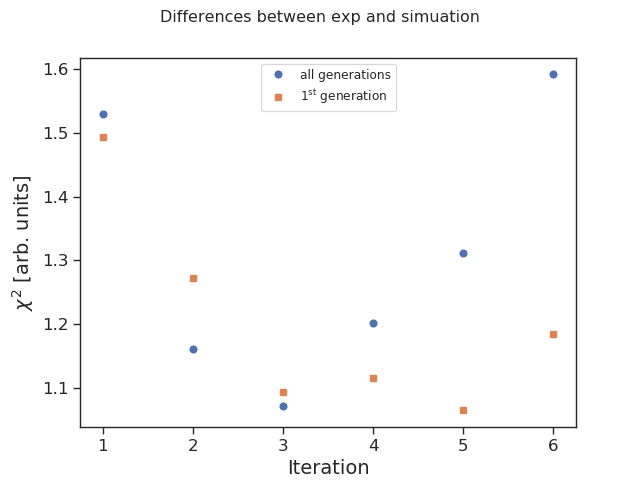}
	\caption{$\chi^2$ between the synthetic and experimental coincidence and first generation data for each iteration.}
	\label{fig:chi2}
\end{figure}

In Fig. \ref{fig:spectra} we compare the experimental coincidence data with the synthetic data from different iterations. All spectra have been normalized to obtain the probability $P(E_\gamma)$ for the emission of a $\gamma$-ray with energy $E_\gamma$ in the decay cascade from a level in the excitation energy bin $E_x$. This removes any dependence on the simulated vs. measured number of $\gamma$-rays and of a potential mismatch of the population cross-section as a function of the excitation energy $E_\mathrm{x}$. The $\chi^2$ differences over whole extraction region (see Sec. \ref{sec:exp}) are displayed for each iteration in Fig. \ref{fig:chi2}. We find that iteration 3 improves the reproduction of the experimental coincidence spectra by about 50\%, compared to the initial analysis, iteration 1. Higher iterations give a reasonable reproduction of the first generation spectra, but show an increased deviation of the (all generations) coincidence spectra. This might be explained by an overcompensation for $E_\gamma>4$ MeV as discussed above. Additionally, a closer analysis of the first vs all generations spectra indicate a too high probability to decay through a specific state, or set of states, with $E_x\approx1.3$ MeV. This is already visible for iteration 3 in Fig. \ref{fig:spectra}, but the mismatch increases for the higher iterations.

In the described procedure, we used a Monte-Carlo approach to simulate the nucleus and its behavior, therefore, the results may vary between different realizations from the same input parameters. However, we found that in the case of a heavy nucleus the level density was so high that the effects could be neglected for this analysis.

\begin{figure}[tb]
	\centering
	\includegraphics[width=1\linewidth]{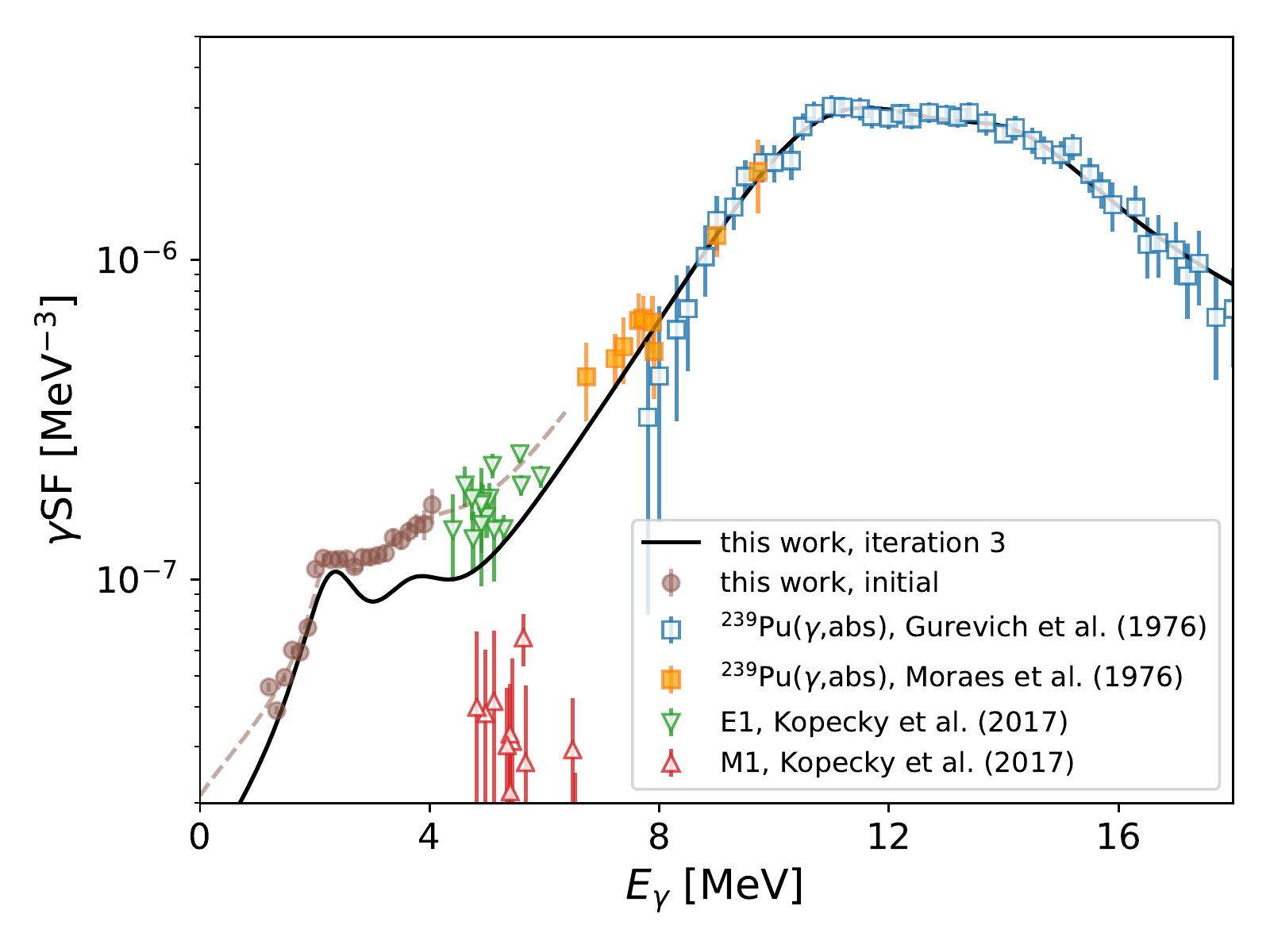}
	\caption{Proposed $\gamma$SF (iteration 3) compared to the initial analysis (see Sec. \ref{sec:exp_extraction}) and measurements by Kopecky \textit{et al.} \cite{Kopecky2017, Chrien1985}, \citet{Moraes1993} and \citet{Gurevich1976}.}
	\label{fig:gsf_final}
\end{figure}

\section{Discussion}

As noted in Ref. \cite{Zeiser2018a}, the Oslo Method does not intrinsically account for differences in the spin-parity distributions $g_\mathrm{pop}$ and $g_\mathrm{int}$; when there is a significant spin-parity mismatch the resulting \textit{apparent} NLD and $\gamma$SF will be distorted compared to the \textit{a priori} \textit{true} NLD and $\gamma$SF. This effect can be observed in Fig. \ref{fig:nld_gsf_RAINIER} by comparing the input for iteration 3 to RAINIER to the results after application of the Oslo Method. The presented method takes into account $g_\mathrm{pop}$ and $g_\mathrm{int}$ and generates synthetic coincidence datasets. As the \textit{apparent} NLD and $\gamma$SF extracted with the Oslo Method from synthetic and experimental coincidences data suffer the same distortions, we can identify a consistent set of NLD and $\gamma$SF from those simulations that lead to an \textit{apparent} NLD and $\gamma$SF that match the results from the experimentally obtained coincidences (Sec. \ref{sec:exp_extraction}). In Fig. \ref{fig:nld_gsf_RAINIER} it can be observed that this is the case for the input NLD and $\gamma$SF to the $3^\mathrm{rd}$ iteration. Future studies are recommended to establish the sensitivity of the current approach. It is for example possible to find other suitable decompositions and other empirical models to describe the $\gamma$SF in the fitting procedure in Sec. \ref{sec:method}. This would effect the $\gamma$SF and NLD derived from this method. Ideally, one could couple the RAINIER simulations with a Monte Carlo Markov chain code \cite{Gelfand1990, Robert2011} directly (without iterating through the Oslo Method) and find the posterior probability of different NLD and $\gamma$SF combinations to match the experimental observations. However, for a heavy nucleus such as $^{240}\mathrm{Pu}$ each iteration takes about 50 h on a single core Intel E5-2683v4 2.1 GHz, such that the computational costs quickly render a full-scale parameter search unfeasible.

In Fig. \ref{fig:gsf_final}, we compare the input $\gamma$SF of the {$3^\mathrm{rd}$ } iteration to the result of the initial analysis (see Sec. \ref{sec:exp_extraction}) and the measurements of \citet{Kopecky2017}, \citet{Moraes1993} and \citet{Gurevich1976}. The absolute scale of the proposed $\gamma$SF is lower than in the initial analysis, which is attributed to the increased NLD (see Fig. \ref{fig:nld_gsf_RAINIER}, left panels), as can be seen from Eq. \eqref{eq:GammaGamma}.

Around $6~\mathrm{MeV}$, the derived $\gamma$SF is significantly lower than the measurements by Kopecky \textit{et al.} \cite{Kopecky2017, Chrien1985}. However, there are two ways to resolve the apparent discrepancy: First, according to the original analysis \cite{Chrien1985}, the data of Kopecky \textit{et al.} \cite{Kopecky2017, Chrien1985} have a systematic normalization uncertainty of 30\% (only the statistical errors are plotted). Second, our results have little sensitivity to the $\gamma$SF above approximately $4~\mathrm{MeV}$. Thus, we could add another resonance at $\approx 6-8~\mathrm{MeV}$ without changing any other observables, like the shape of the extracted $\gamma$SF or $\langle \Gamma_\gamma\rangle$.

The retrieved $\gamma$SF reveals an excess strength between $2-4~\mathrm{MeV}$ on the hypothetically smooth tail of the GDR. This is consistent with the location of the low-energy orbital M1 SR \cite{Heyde2010}. Several other studies in the actinide region using (d,p) reactions and the Oslo Method have observed a similar excess~\cite{Guttormsen2012, Guttormsen2013,Guttormsen2014, Tornyi2014a, Laplace2015}. However, we expect that the spin-parity distributions may also have biased the NLD and $\gamma$SF obtained in those experiments, and therefore plan to reanalyze the extracted strength with the present method.

\section{Conclusions}
\label{sec:con}

We have developed an iterative procedure to correct for the bias introduced in the Oslo Method when the spin-parity dependent population probability $g_\mathrm{pop}$ differs significantly from the spin-parity distribution of the intrinsic levels $g_\mathrm{int}$. We have calculated $g_\mathrm{pop}$ for the $^{239}\mathrm{Pu}(\mathrm{d},\mathrm{p})^{240}\mathrm{Pu}$ experiment performed at the OCL within the Green's Function Transfer formalism. Using the nuclear decay code RAINIER, we have simultaneously retrieved a NLD and $\gamma$SF of $^{240}\mathrm{Pu}$ which are consistent with the experimental analysis. The $\gamma$SF reveals excess strength between $2-4~\mathrm{MeV}$, which can be identified as the orbital M1 SR. The results have been compared to other measurements and the origin of the differences has been addressed.

\acknowledgements
We would like to thank J.~C.~Müller, E.~A.~Olsen, A.~Semchenkov, and J.~C.~Wikne at the Oslo Cyclotron Laboratory for providing the stable and high-quality deuteron beam during the experiment. We are indebted to R.~Capote for discussions on the implementation of the OMP and L.E.~Kirsch for his help  explaining and extending the of capabilities of RAINIER. 
This work was supported by the Research Council of Norway under project Grants No. 263030 and 262952, and by the National Research Foundation of South Africa. A.C.L. gratefully acknowledges funding from the European Research Council, ERC-STG-2014 Grant Agreement No. 637686. 
We gratefully acknowledge support of the U.S. Department of Energy by Lawrence Livermore National Laboratory under Contract No. DE-AC52-07NA27344 and by the Lawrence Berkeley National Laboratory under Contract No. DE-AC02-05CH11231. This work was supported by the U.S. Department of Energy (DOE), National Nuclear Security Administration (NNSA) under Awards DE-NA0002905 and DE-NA0003180, the latter via the Office of Defense Nuclear Nonproliferation Research and Development (DNN R\&D) through the Nuclear Science and Security Consortium. We acknowledge support by DOE Office of Science, Office of Nuclear Physics, under the FRIB Theory Alliance award DE-SC0013617. This work was supported by the National Research Foundation of South Africa Grant No. 118846.

\bibliography{../../bib}              

\end{document}